\documentclass[fleqn,twoside]{article}
\usepackage{amssymb}
\usepackage{espcrc2}

\usepackage{graphicx}
\usepackage[figuresright]{rotating}

\newcommand{\err}[2]{${\scriptstyle {}^{+{#1}}_{-{#2}}}$}

\newcommand{\AmS}{{\protect\the\textfont2
  A\kern-.1667em\lower.5ex\hbox{M}\kern-.125emS}}

\title{
\vspace{-1.0cm}
\hfill\begin{minipage}{0pt}\scriptsize \begin{tabbing}
        \hspace*{\fill} Edinburgh 2001/16\\ \end{tabbing}\end{minipage}\\[8pt]
Light hadron spectrum using an O(a)-improved Wilson action with two dynamical quark flavours }

\author{Derek J. Hepburn\address[MCSD]{Department of Physics and Astronomy,
        University of Edinburgh, EH9 3JZ, Scotland.},\\UKQCD Collaboration.}
       
\begin{document}

\begin{abstract}
  I present results of recent work for the UKQCD collaboration on the
  light hadron spectrum using a non-perturbatively O(a)-improved
  Wilson action with two degenerate flavours of dynamical quarks on a
  $16^3\times32$ lattice. Values of the bare gauge coupling and bare
  dynamical quark mass were chosen to keep the lattice spacing fixed
  at two distinct values in physical units as determined through the
  Sommer scale parameter, $r_{0}$. I also include results for a
  quenched ensemble with a lattice spacing matching one of these
  dynamical sets and results from an exploratory run at lighter bare
  quark mass.  \vspace{1pc}
\end{abstract}

\maketitle

\section{Introduction}
The prediction of the spectrum of light hadron masses is one of the
fundamental aims of lattice QCD. This has already been done
extensively in the quenched approximation by several collaborations.
The results for the quenched spectrum have been found to lie within
10\% of the experimental values. This is encouraging and motivates a
full calculation with dynamical quarks. The results of such a study
would test the appropriateness of QCD as the theory of the strong
interaction.

In this paper I present work on the light hadron spectrum for the
UKQCD Collaboration using a non-perturbatively O(a)-improved Wilson
action with two degenerate dynamical quarks. Much of this work can be
found in a recent paper on unquenching effects by the UKQCD
Collaboration \cite{csw202}. The three dynamical simulations and one
quenched simulation used to study unquenching were performed at values
of $\beta$ and $\kappa_{sea}$(where appropriate) falling on a
trajectory of fixed lattice spacing. This was done to minimise
variations due to residual discretisation errors and finite volume
effects. In addition to these four simulations I include results from
two dynamical simulations lying on a different fixed lattice spacing
trajectory and one exploratory run at lighter quark mass to test the
limits of the current algorithms. The fixed lattice spacing
trajectories were determined using the matching techniques in
\cite{matching} and the scale set using $r_{0}$ in all simulations.

\section{Simulation Details}

\begin{table*}[tb]

\caption{
  Summary of simulation details, statistics and specific observables
  with statistical errors only. Gauge configuration for the
  (5.2,0.13565) dataset have been separated by 20 trajectory steps (
  as opposed to 40 for the other datasets) since statistics were low.
  }

\label{sim_details}

\begin{tabular}{llllllll}
  \hline
  $\beta$ & $\kappa_{sea}$ & $c_{sw}$ & \#conf. & $a[fm]$ & $L/r_{0}$ & $M_{PS}/M_{V}$ & $\kappa_{val}$\\
  \hline
  5.20 & 0.13565 & 2.0171 & 141 & 0.0941(8) & 3.07(3) & 0.532(16) & 0.13565\\
  5.20 & 0.1355  & 2.0171 & 208 & 0.0972(8) & 3.17(3) & 0.58(2) & 0.1340 \ 0.1345 \ 0.1350 \ 0.1355\\
  5.25 & 0.1352  & 1.9603 & 206 & 0.0953(9) & 3.12(3) & 0.705(8) & 0.1342 \ 0.1347 \ 0.1352 \ 0.1357\\
  5.20 & 0.1350 & 2.0171 & 151 & 0.1031(09) & 3.37(3) & 0.700(12) & 0.1335 \ 0.1340 \ 0.1345 \ 0.1350\\
  5.26 & 0.1345 & 1.9497 & 102 & 0.1041(12) & 3.40(4) & 0.783(5) & 0.1335 \ 0.1340 \ 0.1345 \ 0.1350\\
  5.29 & 0.1340 & 1.9192 & 101 & 0.1018(10) & 3.32(3) & 0.835(7) & 0.1335 \ 0.1340 \ 0.1345 \ 0.1350\\
  5.93 & - & 1.82   & 623 & 0.1040(03) & 3.39(1) & - & 0.1327 \ 0.1332 \ 0.1334 \ 0.1337\\
  &&&&&&& 0.1339\\
  \hline

\end{tabular}

\end{table*}

All of the simulations discussed here were performed on a
$16^{3}\times32$ lattice. For the dynamical simulations we used the
Wilson gauge field action together with the Sheikholeslami-Wohlert
O(a)-improved Wilson gauge-fermion action \cite{SW_imp}. The gauge
configurations were generated using the Hybrid Monte Carlo algorithm
\cite{hmc}.

Quark propagators were calculated using the O(a)-improved
gauge-fermion term for both dynamical and quenched simulations. We
used the final published values of the O(a)-improvement coefficient
$c_{sw}$, as determined by the Alpha collaboration \cite{alpha_csw},
in the generation of dynamical gauge fields and quark propagators.

In what follows I distinguish the values of the hopping parameter,
$\kappa$, used in the generation of gauge configurations and quark
propagators by $\kappa_{sea}$ and $\kappa_{val}$ respectively. The
cases where $\kappa_{val}\neq\kappa_{sea}$ were studied to allow a
partially quenched analysis to be performed.

I summarise the simulation parameters in Table 1. The value of
$L/r_{0}$ is included as an indicator of the onset of finite size
effects. The parameter, $r_{0}$, was determined from calculations of
the inter-quark potential and was used to set the scale. I also
include $M_{PS}/M_{V}$ to indicate how light the simulations are. This
ratio was calculated at $\kappa_{sea}=\kappa_{val}$.

It can be seen that all the simulations were performed
with relatively coarse lattices. This was to avoid the appearance of
finite size effects and was limited by the available computer
resources. Previous analysis by UKQCD \cite{finite} has shown that
finite size effects in the light hadron spectrum are negligible
provided

\begin{equation}
L/r_{0} \gtrsim 3.2\ .
\end{equation}

\section{Fitting Hadron Correlators}

In this work I have concentrated on determining the masses of the
pseudoscalar and vector mesons and the nucleon and delta baryons. The
hadronic interpolating operators used in the calculation of the
necessary two-point correlation functions were either local or smeared
using the fuzzing procedure of [7]. I use a nomenclature where an FL
correlator is one that is fuzzed at sink and local at source.

It should be noted that hadron correlators were calculated for all the
possible combinations $\kappa_{val}$ as given in Table
\ref{sim_details} although baryons with non-degenerate constituent
quark masses have not been analysed yet.

The meson masses were determined from their respective correlation
functions by correlated $\chi^{2}$ minimisation with the following
ansatz for correlators at time, $t$,

\begin{equation}
\label{ansatz}
C(t) = \sum_{n=0}^{N} A_{n} \left( e^{-M_{n}t} + e^{-M_{n}(T-t)} \right).
\end{equation}
Here $T$ is the temporal extent of the lattice, $A_{n}$ is a time
independent factor and $M_{n}$ is the mass of the $n^{th}$ excited
state with the ground state($n=0$) being the mass we wish to
calculate. For baryons a similar ansatz was used but with the second
exponential term omitted.

\begin{figure}[ht]
\includegraphics*[width=7.5cm,height=5.5cm]{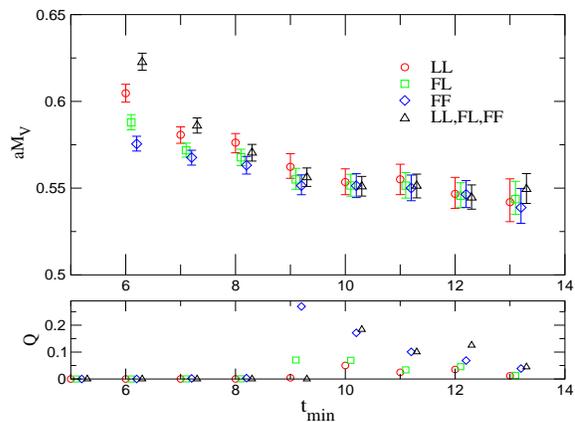}
\caption{Vector fit masses for the $\beta=5.25, \kappa_{sea}=0.1352$
dataset with $\kappa_{val}=\kappa_{sea}$ versus the time range minimum
with $t_{max}$ fixed at $15$.}
\label{sliding}
\end{figure}

\begin{table*}[htb]
\caption{Preliminary hadron mass results in lattice units for the
$(5.2,0.13565)$ and $(5.25,0.1352)$ datasets with statistical errors
only. The values for $\kappa_{val3}$ are only relevant for baryonic
results.}
\label{fit_results}

\begin{tabular}{lllllllll}
\hline
$\beta$ & $\kappa_{sea}$ & $\kappa_{val1}$ & $\kappa_{val2}$ & $ aM_{PS}$ & $aM_{V}$ & $\kappa_{val3}$ & $aM_{N}$ & $aM_{\Delta}$\\
\hline
5.20 & 0.13565 & 0.13565 & 0.13565 & 0.253\err{3}{3}& 0.475\err{13}{13} & 0.13565 & 0.66\err{2}{2}& 0.87\err{3}{3}\\
5.25 & 0.1352  & 0.1342 & 0.1342 & 0.497\err{2}{2} & 0.625\err{4}{3} & 0.1342 & 0.934\err{11}{13}& 1.024\err{14}{14}\\
5.25 & 0.1352  & 0.1347 & 0.1342 & 0.472\err{2}{2} & 0.607\err{4}{4} &-&-&-\\
5.25 & 0.1352  & 0.1347 & 0.1347 & 0.445\err{2}{2} & 0.588\err{4}{4} & 0.1347 & 0.866\err{12}{16}& 0.97\err{2}{2}\\
5.25 & 0.1352  & 0.1352 & 0.1342 & 0.445\err{2}{2} & 0.589\err{5}{4} &-&-&-\\
5.25 & 0.1352  & 0.1352 & 0.1347 & 0.418\err{2}{3} & 0.570\err{5}{5} &-&-&-\\
5.25 & 0.1352  & 0.1352 & 0.1352 & 0.389\err{2}{3} & 0.551\err{6}{6} & 0.1352 & 0.80\err{2}{2}& 0.92\err{3}{3}\\
5.25 & 0.1352  & 0.1357 & 0.1342 & 0.418\err{2}{3} & 0.572\err{6}{5} &-&-&-\\
5.25 & 0.1352  & 0.1357 & 0.1347 & 0.389\err{2}{3} & 0.552\err{6}{6} &-&-&-\\
5.25 & 0.1352  & 0.1357 & 0.1352 & 0.358\err{2}{3} & 0.533\err{7}{7} &-&-&-\\
5.25 & 0.1352  & 0.1357 & 0.1357 & 0.324\err{3}{3} & 0.513\err{9}{9} & 0.1357 & 0.74\err{2}{2}& 0.88\err{3}{3}\\
\hline

\end{tabular}

\end{table*}

In addition to the fits to single correlators, simultaneous fits were
performed to LL, FL and FF correlators. Not only did this allow
fitting to a common mass but the amplitude factors, $A_{n}$, for the
three correlators could be written in terms of two common factors,
further constraining the fit and reducing the total number of
parameters.

The hadron mass analysis proceeded in all cases by performing fits
over a series of time ranges. As an example Figure \ref{sliding} shows
the dependence of the fitted ground state mass for a vector meson on
the minimum time of the fit range, $t_{min}$. Final fit ranges were
chosen when the goodness of fit parameter, $Q$, indicated the ansatz
was appropriate and when some stability in the fit parameters with
respect to small changes in the fit range had been reached. The
statistical errors on the fit parameters were calculated by the
bootstrap method.

\section{Results}

The fit results for all but the $(\beta=5.25, \kappa_{sea}=0.1352)$
and $(\beta=5.2, \kappa_{sea}=0.13565)$ datasets can be found in
\cite{csw202}. I present preliminary results for these two remaining
datasets in Table \ref{fit_results}. Fits to mesonic correlators were
performed for all combinations of the valence quark masses, as given
by $\kappa_{val1}$ and $\kappa_{val2}$. Baryonic fits have only been
performed with degenerate valence quarks.

All of these results are from simultaneous fits to LL, FL and FF
correlators. The results for the $(5.2,0.13565)$ dataset and the
mesonic fits for the $(5.25,0.1352)$ dataset were produced with a
ground state only fit ansatz (i.e. $N=0$ in equation \ref{ansatz}).
The remaining baryonic fits were performed with the first excited
state contribution included.

\section{Acknowledgements}

I would like to thank the Carnegie Trust for the Universities of
Scotland and the European Community's Human potential programme under
HPRN-CT-2000-00145 Hadrons/LatticeQCD for their financial support of
this work.

\end{document}